\documentclass[twocolumn,showpacs,showkeys,preprintnumbers,amsmath,amssymb,prl]{revtex4}

\usepackage{cancel}
\usepackage{epsfig}
\usepackage{mathrsfs}
\usepackage{graphicx}
\usepackage{dcolumn}
\usepackage{bm}
\usepackage{fancyhdr} 
\usepackage{verbatim}
\usepackage[all, knot]{xy}
\xyoption{arc}
\usepackage{amsmath}
\usepackage{amssymb}
\usepackage{revsymb}
\usepackage{slashed}

\usepackage[usenames]{xcolor}

\definecolor{red_ud}{rgb}{1,0,0} 
\definecolor{red_cs}{rgb}{.75,0,0} 
\definecolor{red_tb}{rgb}{.5,0,0} 

\definecolor{blue_ud}{rgb}{0,0,1} 
\definecolor{blue_cs}{rgb}{0,0,.75} 
\definecolor{blue_tb}{rgb}{0,0,.5} 

\definecolor{green_ud}{rgb}{0,1,0} 
\definecolor{green_cs}{rgb}{0,.75,0} 
\definecolor{green_tb}{rgb}{0,.5,0} 

\definecolor{lilac}{rgb}{.78, .05, .78}

\usepackage{subfigure} 

\begin{document}


\title{Quantum lattice gas model of Fermi systems with relativistic energy relations}

\author{Jeffrey Yepez}
\date{June 5, 2013}

\address{Air Force Research Laboratory/Directed Energy, Air Force Maui Optical \& Supercomputing Observatory, Kihei, Hawai`i  96753\\
and the Department of Physics and Astronomy, University of  Hawai`i at Manoa\\
Watanabe Hall, 2505 Correa Road, Honolulu, Hawai`i 96822
 }

\begin{abstract}
Presented are several example quantum computing representations of quantum systems with a relativistic energy relation.  Basic unitary  representations  of free Dirac particles and BCS superconductivity are given.  Then, these are combined into a novel unitary representation of a Fermi condensate superfluid.  The modeling approach employs an operator splitting method that is an analytically closed-form product decomposition of the unitary evolution operator, applied in the high-energy limit.  This allows the relativistic wave equations to be cast as unitary finite-difference equations. The split evolution operators (comprising separate kinetic  and interaction energy evolution terms) serve as quantum lattice gas models  useful for  efficient quantum simulation. 
\end{abstract}



\maketitle

{\it Introduction.}---In this Letter we consider a 
 representation of the particle dynamics where the quantum probability amplitude fields that represent the particle configurations in spacetime are taken to be discrete fields. Normally, quantum field theory is considered to be an effective  low-energy or low-momentum theory from some unknown high-energy theory, where ``low" is  defined relative to a momentum cutoff scale.  The  cutoff is a pragmatic device  requiring no knowledge about the high-energy theory, for example, of a possible lattice-like structure that is resolved at the small spatial scale.  Yet, tacitly  acknowledging a grid scale without exploiting its structure has a pitfall:  low-energy quantum field theory is not exactly computable. Here we consider a representation that is exactly computable.  Its application to quantum field theories of many particles requires  an implementation on a scalable quantum computer. Yet, for its analytical specification, the representation is explained for a  Dirac particle and  a self-consistent Fermi superfluid.   The representation is a high-energy model that is a digital version of relativistic quantum dynamics.  

The foundational tenet 
of the approach presented herein is the existence of large energy and momentum cutoffs that, in  turn, imply the existence of a spacetime grid.  Another  tenet is that  quantum particle dynamics is exactly computable on a spacetime grid.   One constructs a finite-dimensional unitary quantum algorithm applied at the large cutoff scale. This approach employs an ordered collection of qubits and quantum gates acting on qubit pairs.   
 The quantum nature of the resulting particle dynamics is attributed to two  physical degrees of freedom: (1) the number of qubits used to encode the informational state at each grid point (which is a fixed number for a particular model) and (2) a quantum knot invariant of the quantum network \cite{PhysRevA.81.022328}  determined by the type of hermitian number operators used to generate the particle motion and particle-particle interaction. This is  a quantum lattice gas model, for example useful for simulating  superfluid dynamics with ideal scaling on supercomputers  \cite{yepez:084501}. 
Quantum lattice gases were one of the earliest quantum algorithms devised \cite{riazanov-spj58,feynman-65-1st-qlga,JSP.53.323,FPL.10.105,yepez-afosr96,yepez_96_tech_report,PhysRevD.49.6920,PhysRevA.54.1106,PhysRevE.55.5261,PhysRevE.57.54}.  Feynman's original quantum lattice gas algorithm dates all the way back to 1946 \cite{feynman-cit46},  and is known as the Feynman chessboard model  \cite{jacobson-jpamg84}\footnote{The quantum lattice gas representation constitutes a universal model of quantum computation.
}.

In computational physics representations,  a Trotter decomposition of the unitary evolution operator is often used.  The Lie-Trotter product formula separates the kinetic energy  $h_\circ$ and interaction potential $h'$  during the time evolution, $e^{-it(h_\circ + h')} = \lim_{n\rightarrow\infty} \big(e^{-it \,h_\circ/n}e^{-it \, h'/n}\big)^n$ \cite{Trotter_JSTOR_1959}. This is done, for example,  in the path integral representation of quantum mechanics \cite{RevModPhys.20.367}.   However, the limit is not exactly computable.  Here we introduce a novel formula $e^{-i \arccos\! \sqrt{1-E^2\tau^2} \,(h_\circ + h')/E}= e^{ i \ell h_\circ} e^{-i  \arccos\!\sqrt{1-m^2\tau^2}\,{h'}/E}$ that is exactly computable as it does not require a limiting procedure. Here $\ell$ and $\tau$ denote the grid length and grid time,  and $E$ denotes the energy scale of particle dynamics.    Such a closed-form decomposition is useful, for example, for quantum computational representations of quantum field theories with a relativistic energy relation.

 There are  a few fundamental types of number operators.  If we denote the joint number operator by $N$, then it is either: 
(I) an involution operator $N^2=\pm 1$, (II) an idempotent operator $N^2= \pm N$, or (III) a tri-idempotent operator $N^3=\pm N$ that is neither involution nor idempotent.    So the  number operator's type is specified by the order of its infolding (whether involution, idempotent, tri-idempotent) and whether regular or skew ($\pm 1$).

Writing the unitary evolution operator as $U = e^{- i  \vartheta N}$ where $\vartheta$ is a real-valued gate angle having units of radians, $N$ is dimensionless.  The reduced Planck constant $\hbar \equiv m_\circ \ell^2/(2\pi \tau)$ may be  parametrized by  mass unit $m_\circ$. This energy scale parametrizes the dynamics through the gate angle $E \tau/\hbar= 2 \pi E/(m_\circ \ell^2/\tau^2)$.  Furthermore, with the given energy scale, one specifies the  computational Hamiltonian as $H \equiv E N$.
The three types of regular unitary  operators,  serving as the building blocks of a quantum lattice gas model,  can be parametrized to express $U=1-i\tau H/\hbar+\cdots$ with favorable numerical accuracy:  
\begin{itemize}
\itemsep=0pt
  \item $
  U_0 \! =\!  
  e^{- i   \arccos\!\sqrt{1-(\frac{E\tau}{\hbar})^2} \left(\frac{H}{E}\right)} = 1+ \epsilon\!\left(\frac{E\tau}{\hbar}\right) - i \frac{\tau H}{\hbar}
  $
  
  \item $U_1 \!= \! 
  e^{- i   \arccos\!\sqrt{1-(\frac{E\tau}{\hbar})^2} \left(\frac{H}{E}\right)} = 1+ \epsilon\!\left(\frac{E\tau}{\hbar}\right) N- i \frac{\tau H}{\hbar}$
  \item $U_2 \!= \! 
  e^{- i   \arccos\!\sqrt{1-(\frac{E\tau}{\hbar})^2} \left(\frac{H}{E}\right)} = 1+ \epsilon\!\left(\frac{E\tau}{\hbar}\right) N^2- i \frac{\tau H}{\hbar}$,
\end{itemize}
where
\(
\epsilon(x)\equiv  \sqrt{1-x^2}-1.
\)
The involution case is  special  because the computational Hamiltonian is exactly the desired physical Hamiltonian with the $\epsilon$ term only causing an overall phase shift.  We can exploit this property, for example, when constructing a quantum lattice gas  to model the  quantum dynamics of a system with a relativistic energy relation.  Three  examples follow.

{\it Free Dirac particles.}---Let us consider an example applicable to modeling Dirac particles where the grid-level quantum evolution equation is a product of two unitary operators, respectively generated by a kinetic stream operator $N_\circ\equiv h_\circ/(p\ell/\tau)$ and a chirality breaking operator $N'\equiv h'/E$, where $p$ is   momentum.  
According to the Baker-Campbell-Hausdorff (BCH) formula for noncommuting  generators (${N_\circ} {N'} - {N'} {N_\circ}  \ne 0$), one  expects  the  evolution operator formally written as $e^N$ (and formed as a product of $e^{N_\circ}$ and $e^{N'}$) to have  a Hamiltonian  with a cascading structure   of nested commutators:
  \(
  e^N = e^{N_\circ} e^{N'} 
  \) with
\begin{equation}
   N = {\sum_{n>0}\frac{(-1)^{n-1}}{n}\sum_{\stackrel{r_i+s_i> 0}{1\le i \le n}} \frac{{N_\circ}^{r_1}{N'}^{s_1} \cdots {N_\circ}^{r_n}{N'}^{s_n}}{r_1! s_1! \dots r_n! s_n!}},
\end{equation}
  where for simplicity we neglect writing down the gate angles.  
  From an analytical perspective, the BCH formula is a catastrophe.  Either the evolution operator or the Hamiltonian generator can be expressed compactly in closed form, but both cannot be simultaneously expressed this way.
  Yet, in a quantum lattice gas method presented herein, we can avoid the BCH catastrophe by restricting the generators ${N_\circ}$ and ${N'}$ to be either an involution, idempotent, or tri-idempotent operator.

In the simplest case, both  number operators would be involution operators.  So, for example, we could have
  \begin{subequations}
\begin{eqnarray}
  e^{i  p\ell {N_\circ}} &= &\cos\!p  \ell +  i N_\circ \sin\!p\ell, 
 \\
 e^{-i \arccos\!\sqrt{1-m^2\tau^2}\,{N'}} & =&   \sqrt{1-m^2\tau^2} - i \,m\tau {N'},
\end{eqnarray} 
 \end{subequations}
for the involution case   ${N_\circ}^2 = 1$ and ${N'}^2 =1$, 
where the magnitude of the particle's momentum is $p\equiv |\bm{p}|$ and its mass is $m$. 
Suppose further that we choose $N_\circ \equiv \bm{\alpha}\cdot \hat{\bm{p}}$  and $N' \equiv \beta \,e^{i  \bm{\alpha}\cdot \bm{p}\ell } $, where $\bm{\alpha}$ and $\beta$ are Dirac matrices and where $\hat{\bm{p}} \equiv \bm{p}/|\bm{p}|$. 
Then, the unitary evolution equation at the temporal grid scale is $\psi(t+\tau, \bm{x}) = U_\text{\tiny D}^\text{\tiny H.E.}\psi(t, \bm{x})$, and the product evolution operator is
 \begin{subequations}
 \label{high_energy_Dirac_product_evolution}
 \begin{eqnarray}
 \label{high_energy_Dirac_product_evolution_a}
U_\text{\tiny D}^\text{\tiny H.E.}
&\equiv &
e^{ i p\ell {N_\circ}} e^{-i  \arccos\!\sqrt{1-m^2\tau^2}\,{N'}}
\\
 \label{high_energy_Dirac_product_evolution_b}
 &=&
e^{i  \bm{\alpha}\cdot \bm{p}\ell }    \sqrt{1-m^2\tau^2} - i \,m\tau \,e^{i  \bm{\alpha}\cdot \bm{p}\ell }  \beta \,e^{i  \bm{\alpha}\cdot \bm{p}\ell } 
,
\quad
\end{eqnarray} 
 \end{subequations}
which is the high energy {\scriptsize (H.E.)} representation of the Dirac particle dynamics. 
Since the Dirac matrices anticommute, the second term on the righthand side  of (\ref{high_energy_Dirac_product_evolution_b}) simplifies to $- i \,m\tau \,e^{i  \bm{\alpha}\cdot \bm{p}\ell }  \,e^{-i  \bm{\alpha}\cdot \bm{p}\ell }  \beta =-i m \tau \beta$.  The first term on the righthand side of (\ref{high_energy_Dirac_product_evolution_b}) can be rewritten using the relativistic energy relation $E^2 = p^2 + m^2$ (in natural units $\hbar=1$ and $c=\ell/\tau=1$) that,  upon multiplying through by the grid scale squared and adding one to both sides, may be expressed as
\(
1-(m\tau)^2 = 1- (E \tau)^2 + (p\ell)^2.
\)
We may take the ``square root'' of this equation to be
\begin{equation}
\sqrt{1-(m\tau)^2}  \,e^{i  \bm{\alpha}\cdot \bm{p}\ell }   = \sqrt{1- (E \tau)^2 }- i \bm{\alpha}\cdot \bm{p} \ell,
\end{equation}
where we took the liberty to choose  otherwise undetermined phases \footnote{It is straightforward to verify that multiplying this square root equation by its  conjugate yields the relativistic energy relation since $\bm{\alpha}^\dagger= \bm{\alpha}$ and $(\bm{\alpha}\cdot\bm{p})^2=p^2$.
}.  
Thus,  (\ref{high_energy_Dirac_product_evolution_b}) may be  written as
  \begin{subequations}
\begin{eqnarray}
U_\text{\tiny D}^\text{\tiny H.E.}
&=&\sqrt{1- (E \tau)^2 }- i\bm{\alpha}\cdot \bm{p} \ell -i \beta\,m \tau
\\
&=&\sqrt{1- (E \tau)^2 }- i E \tau\, \frac{h_\text{\tiny D}}{E}
\\
&=&
\exp\Big[{-i\frac{\arccos\sqrt{1- (E \tau)^2 }}{E} \, h_\text{\tiny D}}\Big]
,
\end{eqnarray} 
 \end{subequations}
where the Dirac Hamiltonian $h_\text{\tiny D} \equiv \bm{\alpha}\cdot \bm{p}  \ell/\tau+ \beta \,m$ is the generator of the unitary time evolution.  The last expression on the righthand side follows because $(h_\text{\tiny D}/E)^2=1$, which is involution. Taylor expanding the gate angle of the unitary evolution operator about $E=0$ gives
\(
{\arccos\sqrt{1- (E \tau)^2 }}/{E} = \tau + {\tau^3 E^2}/{6}+\cdots,
\)
so in the low-energy limit we have $U_\text{\tiny D}^\text{\tiny L.E.}= e^{-i \tau h_\text{\tiny D} }$.  This is the expected form for an effective quantum evolution operator.  However, expressed in the high-energy limit  (or in other words, accurate to grid-level resolution), the exact unitary evolution operator is $U_\text{\tiny D}^\text{\tiny H.E.}= \sqrt{1- (E \tau)^2 }- i  \tau\, h_\text{\tiny D}$.  
So the equation of motion that  specifies the Dirac particle dynamics on the grid is
\begin{equation}
\label{grid_level_Dirac_equation}
\psi(t + \tau, \bm{x}) = \sqrt{1- (E \tau)^2 }\,\psi(x) - i  \, \left(\bm{\alpha}\cdot \bm{p}  \ell+ \beta \,m\tau\right) \psi(x),
\end{equation}
where $x=(t,\bm{x})$. 
 The dimensionless quantities $\bm{p} \ell$,  $m \tau$, and $E \tau$, satisfying $E\tau =\sqrt{(p\ell)^2 + (m\tau)^2}$,
parametrize an exactly computable dynamics on a grid \cite{yepez_arXiv1106.0739_gr_qc}.

{\it BCS superconductivity.}---As a second example, suppose that an interaction operator is generated by a joint number operator that is tri-idempotent, i.e. for the  case ${N'}^3 = {N'}$ and ${N'}^2 \ne {N'}$.
The case of a tri-idempotent joint number operator applies to BCS superconductivity (in a self-consistent treatment). For a system with $Q$ qubits, the qubit creation and annihilation operators   anticommute, 
\(
\{   a_\alpha,   a^\dagger_\beta\}  =  \delta_{\alpha\beta},
\) 
\(
\{   a_\alpha,   a_\beta\}  =   0,
\) 
\(
\{   a^\dagger_\alpha,   a^\dagger_\beta\}  =  0,
\) 
for integer $\alpha,\beta \in [1,Q]$.  
Using the Jordan-Wigner transformation \cite{JordanWigner1928}, an annihilation operator is decomposed into a tensor product 
\(
a_\alpha=\sigma_z^{\otimes \alpha-1} \otimes\, a\otimes \bm{1}^{\otimes Q-\alpha},
\)
where $a$ is the singleton annihilation operator.  
The number operator $  n_\alpha \equiv   a^\dagger_\alpha   a_\alpha$ has eigenvalues of 1 or 0 when acting  the $\alpha\hbox{th}$ qubit, corresponding to states $|1\rangle_\alpha$ or $|0\rangle_\beta$, respectively. In the BCS case, the joint number operator that generates the nonlinear interaction is
\begin{eqnarray}
\nonumber
{N}_{
\alpha\beta} ^{\pm}
   &=&
    \left.\frac{1}{2}\middle(1\pm\frac{{\cal E}}{E}\right) a^\dagger_\alpha a_\alpha
+
\left.\frac{1}{2}\middle(1\mp\frac{{\cal E}}{E}\right) a_\beta a_\beta^\dagger
\quad
\\
&
\pm
&
    \frac{\Delta}{2 E}
a^\dagger_\alpha a_\beta^\dagger 
\pm
\frac{\Delta^\ast}{2 E}
a_\beta   a_\alpha,
\label{BCS_joint_number_operators}
\end{eqnarray}
where $\alpha$ and $\beta$ denote the two qubits encoding two electrons entangled  as a Cooper pair.  Using the anticommutation relations for the creation and annihilation operators, one can verify that  ${N}_{\alpha\beta}^3={N}_{\alpha\beta}$ and that ${N}_{\alpha\beta}^2\neq {N}_{\alpha\beta}$.   
The BCS Hamiltonian is $H_{\alpha\beta}^\text{\tiny BCS}={E}\, {N}_{\alpha\beta}$. 
Hence, the unitary interaction operator for a BCS superconducting fluid may be written as an entangling quantum gate
  \begin{subequations}
     \label{high_energy_BCS_product_evolution}
\begin{eqnarray}
U^\text{\tiny BCS}_{\alpha\beta}
&=&
\underbrace{ 1 +\epsilon({ E\tau})\left( \frac{H_{\alpha\beta}^\text{\tiny BCS}}{E}\right)^2
- i \tau  {H_{\alpha\beta}^\text{\tiny BCS}}
}_\text{local entangling gate}
\quad
\\
& =&
 e^{-i  \arccos\sqrt{1 -({ E\tau})^2} H_{\alpha\beta}^\text{\tiny BCS} /E},
\end{eqnarray}
 \end{subequations}
where $ \prod_{\langle \alpha\beta \rangle  } U^\text{\tiny BCS}_{\alpha\beta} =  e^{-i  \sum_{\langle \alpha\beta \rangle  }   \arccos\sqrt{1 -({ E \tau})^2} H_{\alpha\beta}^\text{\tiny BCS} /E}$ since entanglement is localized in $\bm{k}$-space. 
The grid-level evolution   is  $\psi(t + \tau, \bm{x}) = U^\text{\tiny BCS}  \psi(x)$, where the quantum state at a point is a 4-spinor.  Both the unitary interaction operator and the hermitian generator are analytically known in closed form. In the low-$E$ limit, the  effective dynamics  is generated by the Bogoliubov-de Gennes (BdG) Hamiltonian  $H^\text{\tiny BdG}_{\alpha\beta}\stackrel{(\ref{BCS_joint_number_operators})}{=}E({N}_{
\alpha\beta} ^{+}-{N}_{\alpha\beta} ^{-})$, acts locally within the Hilbert space spanned by two entangled qubits, and   obeys the BdG equation 
\begin{equation}
\label{BdG_equation_4_spinor_form}
i\hbar\partial_t 
\psi
=
H^\text{\tiny BdG}
\psi,
\quad
\quad
H^\text{\tiny BdG}
=
{\scriptsize
\begin{pmatrix}
      {\cal E}  & 0 & 0 
       &  \Delta        \\
   0 &  0  &  0  & 0
        \\
     0 &  0  & 0 &  0
        \\
     \Delta^\ast  & 0 & 0 
      &  -{\cal E} 
\end{pmatrix}_{\!\!\!\alpha\beta}
}
.
\end{equation}
This effective equation of motion is rather simple  because the BSC superconductivity is associated with  pairwise entanglement between   $|00\rangle_{\alpha\beta}$ and $|11\rangle_{\alpha\beta}$ states. The many-body system is a Fermi condensate superfluid. 

The effective equation of motion (\ref{BdG_equation_4_spinor_form}), in the low-energy limit,  is well known.  What is the  equation of motion in the high-$E$ limit?  
Notice that $(H^\text{\tiny BdG}/E)^2=1$, since $E^2=\sqrt{{\cal E}^2 + |\Delta|^2}$.  Hence, the joint number operator $N^\text{\tiny BdG}\equiv H^\text{\tiny BdG}/E$,  an involution operator, can be used to generate the high-energy  particle dynamics on the grid.   The unitary evolution equation at the  grid scale is $\psi(t+\tau, \bm{x}) = U_\text{\tiny BdG}^\text{\tiny H.E.}\psi(t, \bm{x})$, where  the product evolution operator may be  written as
  \begin{subequations}
   \label{high_energy_BdG_product_evolution}
\begin{eqnarray}
U_\text{\tiny BdG}^\text{\tiny H.E.} 
&=&
 e^{-i \arccos\!\sqrt{1-(E\tau)^2}\, N^\text{\tiny BdG}} 
 \\
 &=&
 \sqrt{1-(E\tau)^2} - i E \tau N^\text{\tiny BdG} 
 \\
 &=&
 \sqrt{1-(E\tau)^2} - i \tau H^\text{\tiny BdG},
\end{eqnarray}
 \end{subequations}
where we employed Euler's identity. 
Therefore, a grid-level representation of the BdG equation  is
\begin{equation}
\label{grid_level_BdG_equation}
\psi(t+\tau, \bm{x}) = \sqrt{1-(E\tau)^2}\, \psi(t, \bm{x}) - i \tau H^\text{\tiny BdG} \psi(t, \bm{x}).
\end{equation}
This representation of the BdG equation illustrates how a number operator, that is an involution, can generate  high-energy particle dynamics.  However, in this example,  $U_\text{\tiny BdG}^\text{\tiny H.E.}$ does not  constitute a quantum algorithm for a Fermi superfluid, per se, because the kinetic energy operator is only parametrized by the eigenvalue ${\cal E}$.  We must go further and find a number operator representation of the  kinetic energy term.  As we did above, we may  use the unitary stream operator $e^{i \bm{\alpha}\cdot \bm{p}\ell}$, applied to each point, as a mechanism to effect the kinetic transport of  the quantum particles in the system.   We illustrate this in the next example.


{\it Nonlinear relativistic Fermi superfluid.}--After applying the quantum informational method to BCS superconductivity, we  now isolate the nonlinear interaction  operator
\begin{equation}
\mathfrak{N}' = \frac{1}{|\Delta|}
\begin{pmatrix}
   0   & \Delta   \\
\Delta^\ast      &   0
\end{pmatrix},
\end{equation}
an involution operator ($\mathfrak{N}'^2=1$) useful for generating a self-consistent pairing interaction.  We may use this to construct a nonlinear  representation of the SU(2) group
\begin{equation}
\Sigma_x \!= \!
{\scriptsize 
\frac{1}{|\Delta|}
\begin{pmatrix}
   0   & \Delta   \\
\Delta^\ast      &   0
\end{pmatrix}
},
\quad
\Sigma_y \!=\!
{\scriptsize 
 \frac{1}{|\Delta|}
\begin{pmatrix}
   0   & -i \Delta   \\
i \Delta^\ast      &   0
\end{pmatrix}
},
\quad
\Sigma_z 
\!=\!
{\scriptsize 
\begin{pmatrix}
   1   & 0   \\
0    &   -1
\end{pmatrix}.
}
\end{equation}
As a generalization of the Dirac gamma matrices in the chiral representation, $\gamma^0=\sigma_x\otimes \bm{1}$ and $\gamma^i = i \,\sigma_y\otimes \sigma_i$, let us consider the nonlinear Dirac gamma matrices 
\begin{equation}
\Gamma^0 = \Sigma_x\otimes \bm{1} 
\qquad 
\text{and}
\qquad
\Gamma^i = i \,\Sigma_y\otimes \sigma_i.
\end{equation}
A nonlinear covariant Lagrangian density is thus
  \begin{subequations}
\label{nonlinear_lagrangian_density}
\begin{equation}
\label{nonlinear_lagrangian_density_a}
{\cal L}^\text{\tiny NL}
=
\hbar c\,
\psi^\dagger
\Gamma^0 
\left(
i \,\Gamma^\mu \partial_\mu
- \frac{|\Delta|}{\hbar c}
\right)
\psi,
\end{equation}
where the reduced Compton wavelength of the effective Dirac particle is $\lambdabar = |\Delta|/(\hbar c)$.   Since $\Sigma_z=\sigma_z$, notice that $\Gamma^0\Gamma^\mu= \gamma^0\gamma^\mu$. So (\ref{nonlinear_lagrangian_density_a})  reduces to
\begin{eqnarray}
\label{nonlinear_lagrangian_density_b}
{\cal L}^\text{\tiny NL}
&=&
i\hbar c\,
\overline{\psi}
\gamma^\mu \partial_\mu
\psi
- 
\overline{\psi}
\gamma^0
\Gamma^0 
|\Delta|
\psi
\\
\label{nonlinear_lagrangian_density_c}
&=&
i\hbar c\,
\overline{\psi}
\gamma^\mu \partial_\mu
\psi
- 
\overline{\psi}
\begin{pmatrix}
    \Delta^\ast   & 0 \\
0      &   \Delta
\end{pmatrix}\otimes \bm{1}
\psi,
\end{eqnarray}
 \end{subequations}
where $\overline{\psi}\equiv\psi^\dagger\gamma^0$.  Thus, the first term in (\ref{nonlinear_lagrangian_density_c})  is  the free part of the Dirac Lagrangian density, ${\cal L}_\circ^\text{\tiny NL}={\cal L}_\circ^\text{\tiny D}$.  With the 4-spinor field $\psi = (\psi_{\text{\tiny L}\uparrow}\;\;\psi_{\text{\tiny L}\downarrow}\;\;\psi_{\text{\tiny R}\uparrow}\;\;\psi_{\text{\tiny R}\downarrow})^\text{\tiny T}$,   the interaction  term in (\ref{nonlinear_lagrangian_density_c}) is explicitly
\begin{equation}
{\cal L}_\text{int}^\text{\tiny NL}
=
\Delta\,\psi^\ast_{\text{\tiny L}\uparrow}
\psi_{\text{\tiny R}\uparrow}
+
\Delta\,\psi^\ast_{\text{\tiny L}\downarrow}
\psi_{\text{\tiny R}\downarrow}
+
\Delta^\ast\,\psi^\ast_{\text{\tiny R}\uparrow}
\psi_{\text{\tiny L}\uparrow}
+
\Delta^\ast\,\psi^\ast_{\text{\tiny R}\downarrow}
\psi_{\text{\tiny L}\downarrow}.
\end{equation}
For an unpolarized Fermi condensate, the pairing potential may be chosen to be
\(
\Delta
=
\lambda
\langle
\psi^\ast_{\text{\tiny R}\uparrow}
\psi_{\text{\tiny L}\uparrow}
\rangle
=
\lambda
\langle
\psi^\ast_{\text{\tiny R}\downarrow}
\psi_{\text{\tiny L}\downarrow}
\rangle,
\)
so we are free to rewrite the interaction  as
\begin{eqnarray}
\nonumber
{\cal L}_\text{int}^\text{\tiny NL}
 &=& 
 \lambda\left(
 \psi^\ast_{\text{\tiny L}\uparrow}\psi_{\text{\tiny R}\uparrow}\,\langle \psi^\ast_{\text{\tiny R}\uparrow} \psi_{\text{\tiny L}\uparrow}\rangle
 +
\langle  \psi^\ast_{\text{\tiny L}\uparrow}\psi_{\text{\tiny R}\uparrow}\rangle\,\psi^\ast_{\text{\tiny R}\downarrow} \psi_{\text{\tiny L}\downarrow}
  \right.
\\
&  + &
\left.
\langle  \psi^\ast_{\text{\tiny L}\downarrow}\psi_{\text{\tiny R}\downarrow}\rangle\,\psi^\ast_{\text{\tiny R}\uparrow} \psi_{\text{\tiny L}\uparrow}
  + 
  \psi^\ast_{\text{\tiny L}\downarrow}\psi_{\text{\tiny R}\downarrow}\,\langle\psi^\ast_{\text{\tiny R}\downarrow} \psi_{\text{\tiny L}\downarrow}\rangle
 \right).
\end{eqnarray}
This nonlinear  two-fermion interaction is a partially contracted (or mean-field) version of the effective four-fermion Nambu and Jona-Lasinio (NJL) interaction
\begin{subequations}
\label{NJL_interaction_lagrangian}
\begin{eqnarray}
{\cal L}^\text{\tiny NJL}_\text{int} &=& \frac{\lambda}{4} \left[\left(\overline{\psi} \psi\right)^2-\left(\overline{\psi}\gamma^5\psi \right)^2\right]
 = \lambda\left(\psi_\text{\tiny L}^\dagger \psi _\text{\tiny R}\middle)\middle(\psi_\text{\tiny R}^ \dagger \psi_\text{\tiny L}\right)
 \qquad\\
 \nonumber
 &=& 
 \lambda\left(
 \psi^\ast_{\text{\tiny L}\uparrow}\psi_{\text{\tiny R}\uparrow}\,\psi^\ast_{\text{\tiny R}\uparrow} \psi_{\text{\tiny L}\uparrow}
 +
  \psi^\ast_{\text{\tiny L}\uparrow}\psi_{\text{\tiny R}\uparrow}\,\psi^\ast_{\text{\tiny R}\downarrow} \psi_{\text{\tiny L}\downarrow}
  \right.
\\
&  + &
\left.
  \psi^\ast_{\text{\tiny L}\downarrow}\psi_{\text{\tiny R}\downarrow}\,\psi^\ast_{\text{\tiny R}\uparrow} \psi_{\text{\tiny L}\uparrow}
  + 
  \psi^\ast_{\text{\tiny L}\downarrow}\psi_{\text{\tiny R}\downarrow}\,\psi^\ast_{\text{\tiny R}\downarrow} \psi_{\text{\tiny L}\downarrow}
 \right),
\end{eqnarray}
\end{subequations}
a well known one flavor model of the superfluid phase of a fermion gas at low temperature \cite{PhysRev.122.345}. 
Thus, we see from a different viewpoint how an unpolarized Fermi condensate in the NJL model  causes chiral symmetry breaking, leading to an effective particle mass of $m = |\Delta|/c^2$.

So to complete our  example,  let us construct a quantum lattice gas representation of (\ref{nonlinear_lagrangian_density}) for a Fermi condensate superfluid.  The first step is to calculate the Hamiltonian for this theory. We do this  by rewriting (\ref{nonlinear_lagrangian_density_c}) as
\begin{subequations}
\begin{eqnarray}
{\cal L}^\text{\tiny NL}
&=&
\psi^\dagger
\left[
i\hbar \partial_t - \gamma^0\bm{\gamma}\cdot \bm{p} \ell/\tau 
- 
\begin{pmatrix}
   0   & \Delta   \\
\Delta^\ast      &   0
\end{pmatrix}\otimes \bm{1}
\right]
\psi
\qquad
\\
&=&
\psi^\dagger
\left(
i\hbar \partial_t - h_\text{\tiny NL}
\right)
\psi.
\end{eqnarray}
 \end{subequations}
Thus, we identify the nonlinear relativistic Hamiltonian as $h_\text{\tiny NL}=\bm{\alpha}\cdot \bm{p} \ell/\tau 
+ 
|\Delta|\, \mathfrak{N}'\otimes \bm{1}$ with $\bm{\alpha}=\gamma^0\bm{\gamma}=\sigma_z\otimes\bm{\sigma}$. 
Following the procedure used above in the free Dirac particle example, we now choose the involution operators $N_\circ \equiv \bm{\alpha}\cdot \hat{\bm{p}}$  and $N' \equiv \mathfrak{N}'\otimes \bm{1} \,e^{i  \bm{\alpha}\cdot \bm{p}\ell }$.
Then, the unitary evolution equation at the  grid scale is $\psi(t+\tau, \bm{x}) = U_\text{\tiny NL}^\text{\tiny H.E.}\psi(t, \bm{x})$, and the product evolution operator is
 \begin{subequations}
 \label{high_energy_Fermi_superfluid_product_evolution}
\begin{eqnarray}
 \label{high_energy_Fermi_superfluid_product_evolution_a}
U_\text{\tiny NL}^\text{\tiny H.E.}
&\equiv&
 e^{ i p\ell {N_\circ}} e^{-i  \arccos\!\sqrt{1-|\Delta|^2\tau^2}\,{N'}} 
\\
\nonumber
&=&
e^{i  \bm{\alpha}\cdot \bm{p}\ell }    \sqrt{1-|\Delta|^2\tau^2} - i \,|\Delta|\tau \,e^{i  \bm{\alpha}\cdot \bm{p}\ell }\,  \mathfrak{N}'\otimes \bm{1} \,e^{i  \bm{\alpha}\cdot \bm{p}\ell } .
\\
 \label{high_energy_Fermi_superfluid_product_evolution_b}
\end{eqnarray} 
 \end{subequations}
This is an exactly computable unitary  representation of the  high-energy Fermi condensate dynamics. 
Making use of the anticommutor relation $\{\sigma_z, \mathfrak{N}'\}=0$, the second term on the righthand side simplifies to $- i \,|\Delta|\tau \,e^{i  \bm{\alpha}\cdot \bm{p}\ell }  \,e^{-i  \bm{\alpha}\cdot \bm{p}\ell }\,  \mathfrak{N}'\otimes \bm{1} =-i |\Delta| \tau \,\mathfrak{N}'\otimes \bm{1}$.  The first term on the righthand side can be rewritten using the BCS relativistic energy relation $E^2 = p^2 + |\Delta| ^2$ (in natural units $\hbar=1$ and $c=\ell/\tau=1$). Now we take the ``square root'' of the relativistic energy relation to be
\begin{equation}
\sqrt{1-(|\Delta| \tau)^2}  \,e^{i  \bm{\alpha}\cdot \bm{p}\ell }   = \sqrt{1- (E \tau)^2 }- i \bm{\alpha}\cdot \bm{p} \ell,
\end{equation}
again taking the liberty to choose  otherwise undetermined phases.
Thus, 
(\ref{high_energy_Fermi_superfluid_product_evolution_b}) may be  written as
  \begin{subequations}
  \label{high_energy_nonlinear_unitary_evolution_operator}
 \begin{eqnarray}
U_\text{\tiny NL}^\text{\tiny H.E.}
&=&
\sqrt{1- (E \tau)^2 }- i\bm{\alpha}\cdot \bm{p} \ell -i |\Delta|  \tau \,\mathfrak{N}'\otimes \bm{1} 
\quad
\\
&=&
\sqrt{1- (E \tau)^2 }- i E \tau\, \frac{h_\text{\tiny NL}}{E}
\\
  \label{high_energy_nonlinear_unitary_evolution_operator_c}
&=&
\exp\Big[{-i\frac{\arccos\sqrt{1- (E \tau)^2 }}{E} \, h_\text{\tiny NL}}\Big]
,
\end{eqnarray} 
 \end{subequations}
where  the nonlinear relativistic Hamiltonian $h_\text{\tiny NL}$ is the generator of the unitary time evolution.  Equation (\ref{high_energy_nonlinear_unitary_evolution_operator_c}) follows because $(h_\text{\tiny NL}/E)^2=1$, which is also  involution.  The unitary grid-level equation of motion is
\begin{eqnarray}
\label{grid_level_Fermi_condensate_superfluid_equation}
\nonumber
\psi(t + \tau, \bm{x}) 
&=&
 \sqrt{1- (E \tau)^2 }\,\psi(t,\bm{x})
\\
&-&  i  \, \left(\bm{\alpha}\cdot \bm{p}  \ell+  \tau\begin{pmatrix}
   0   & \Delta   \\
\Delta^\ast      &   0
\end{pmatrix}\otimes \bm{1} \right) \psi(t,\bm{x}),\qquad
\end{eqnarray}
governing the grid-level Fermi condensate dynamics.

{\it Conclusion.}---The  example models treated in this Letter represent  systems of  fermions with a relativistic energy relation. The dynamics of  Fermi systems can be exactly computed on a grid using product decomposition of the evolution operator, such as (\ref{high_energy_Dirac_product_evolution_a})  
 and  (\ref{high_energy_Fermi_superfluid_product_evolution_a}) that   serve as  quantum algorithms for a many-body system of Dirac particles  and a Fermi superfluid, respectively, and that are efficient on a quantum computer. These quantum algorithms allow us to write down new equation of motions, such as (\ref{grid_level_Dirac_equation}) 
and  (\ref{grid_level_Fermi_condensate_superfluid_equation}) respectively, that are  unitary finite-difference equations with no spatial error terms because of the closed-form decomposition. An accompanying article \cite{PhysRevA_2013_2} presents digital  simulations in 1+1 dimensions employing (\ref{high_energy_Dirac_product_evolution_a}) with external scalar potentials.  Ref.~\cite{PhysRevA_2013_2} explains how to use (\ref{high_energy_Dirac_product_evolution_a}) for digital quantum simulation. When the number of qubits is large, these digital representations  approach the respective usual physical quantum field theories in the low-$E$ limit.  The high-$E$ equations of motions are analytical  deformations of  low-$E$ quantum wave equations. This work validates Feynman's conjecture \cite{feynman-82} regarding exact quantum simulation.   

{\it Acknowledgements.}---This work was supported by grant no. AFOSR 11RV13COR RDSM and the DoD HPCMP  Quantum Computing Program at the MHPCC and conducted under an AFRL-UH Educational Partnership Agreement (2010-AFRL/RD-EPA-03 FY2013 Amendment 1).

\end{document}